\documentclass[12pt]{iopart}
\usepackage{iopams}  
\usepackage{epsfig}
\begin{document}

\title[]{Anomalous waiting times in high-frequency financial data}

\author{Enrico Scalas$^{1,2}$ \footnote{Corresponding author, e-mail: scalas@unipmn.it}, Rudolf Gorenflo$^3$, Hugh Luckock$^4$, Francesco Mainardi$^4$, Maurizio Mantelli$^1$, and Marco Raberto$^4$}

\address{$^1$ Dipartimento di Scienze e Tecnologie Avanzate, Universit\`a del
Piemonte Orientale, Piazza Giorgio Ambrosoli 5, Alessandria, I-15100, Italy}

\address{$^2$ INFM, Unit\`a di Genova, Via Dodecaneso 33, I-16149, Genova, Italy}

\address{$^3$ Erstes Mathematisches Institut, Freie Universit\"at Berlin,
Arnimallee 3, Berlin, D-14195, Germany}

\address{$^4$ School of Mathematics and Statistics, University of Sydney, Sydney,
Australia}

\address{$^5$ Dipartimento di Fisica, Universit\`a di Bologna and INFN Sezione di Bologna, Via Irnerio 46, Bologna, I-40126, Italy}

\address{$^6$ Dipartimento di Ingegneria Biofisica ed Elettronica, Universit\`a di Genova, Via all'Opera Pia 11a, Genova, I-16145, Italy}

\begin{abstract}
In high-frequency financial data not only returns, but also waiting times between consecutive trades are random variables. Therefore, it is possible to apply continuous-time random walks (CTRWs) as phenomenological models of the high-frequency price dynamics. An empirical analysis performed on the 30 DJIA stocks shows that the waiting-time survival probability for high-frequency data is non-exponential. This fact imposes constraints on agent-based models of financial markets. 
\end{abstract}

\pacs{05.40.Jc, 89.65.Gh, 02.50.Ey, 05.45.Tp}

\submitto{Quantitative Finance}

\maketitle

\def\eg{{\it e.g.}\ } \def\ie{{\it i.e.}\ }
\def\sg{\hbox{sign}\,}
\def\sgn{\hbox{sign}\,}
\def\sign{\hbox{sign}\,}
\def\e{\hbox{e}}
\def\exp{\hbox{exp}}
\def\ds{\displaystyle}
\def\dis{\displaystyle}
\def\q{\quad}    \def\qq{\qquad}
\def\lan{\langle}\def\ran{\rangle}
\def\l{\left} \def\r{\right}
\def\lra{\Longleftrightarrow}
\def\arg{\hbox{\rm arg}}
\def\d{\partial}
 \def\dr{\partial r}  \def\dt{\partial t}
\def\dx{\partial x}   \def\dy{\partial y}  \def\dz{\partial z}
\def\rec#1{{1\over{#1}}}
\def\log{\hbox{\rm log}\,}
\def\erf{\hbox{\rm erf}\,}     \def\erfc{\hbox{\rm erfc}\,}
\def\F{\hbox{F}\,}
\def\NN{\hbox{\bf N}}
\def\RR{\hbox{\bf R}}
\def\CC{\hbox{\bf C}}
\def\ZZ{\hbox{\bf Z}}
\def\II{\hbox{\bf I}}

\section{Introduction}

Starting from the second half of the last decade, due to the availability of large financial databases, there has been an increasing interest 
on the statistical properties of high-frequency financial data and on market microstructural properties \cite{goodhart97,ohara99,madhavan00,dacorogna01,raberto01,luckock03}.
Various studies on high-frequency econometrics appeared in the literature and among them
autoregressive conditional duration models \cite{engle97,engle98,bauwens00,lo02}.

The basic remark that in high-frequency financial data not only returns but also waiting times between consecutive trades are random variables \cite{zumbach98} can already be found in previous literature. For instance, it is present in a paper by Lo and McKinlay published in the Journal of Econometrics \cite{lo90}, but it can be traced at least to papers on the application of compound Poisson processes \cite{press67} and subordinated stochastic processes \cite{Clark 73} to finance. Compound Poisson processes have been revisited in the recent wave of interest in high-frequency data modelling \cite{rydberg98,rydberg99,rydberg00}.

Compound Poisson processes belong to the class of continuous-time random walks (CTRWs) \cite{montroll65}, which have been recently applied to finance as well (see Sec. 2 for details). To our knowledge, the application of CTRW to economics dates back, at least, to the 1980s. In 1984, Rudolf Hilfer published a book on the application of  stochastic processes to operational planning, where CTRWs were used for sale forecasts \cite{hilfer84}. The (revisited) CTRW formalism has been applied to the high-frequency price dynamics in financial markets by our research group since 2000, in a series of three papers \cite{scalas00,mainardi00,gorenflo01}. Other scholars have recently used this formalism
\cite{masoliver03a,masoliver03b,kutner03}. However, CTRWs have a famous precursor. In 1903, the PhD thesis of Filip Lundberg presented a model for ruin theory of insurance companies, which was further developed by Cram\'er \cite{lundberg03,cramer30}. The underlying stochastic process of the Lundberg-Cram\'er model is another example of compound Poisson process and thus also of CTRW.

Among other issues, we have studied the independence between log-returns and waiting times for the 30 Dow-Jones-Industrial-Average (DJIA) stocks traded at the New York Stock Exchange in October 1999. For instance, according to a contingency-table analysis performed on General Electric (GE) prices, the null hypothesis of independence can be rejected with a significance level of 1 \% \cite{raberto02}. In this paper, however, the focus is on the empirical distribution of waiting times \cite{scalas05}.

This paper is divided as follows: Sec. 2 is devoted to a summary of CTRW theory as applied in finance; the relation of CTRWs to compound Poisson processes will be presented in some detail. In Sec. 3, following our empirical analysis, the reader can convince him/herself of the main result of this paper: for the 30 DJIA stocks in the period considered (October 1999), the waiting-time survival probability for high-frequency data is non-exponential. Finally, in Sec. 4, a possible explanation of this anomaly will be discussed using exponential mixtures as the analytical tool.

\section{Theory}

The importance of random walks in finance has been known since the seminal
thesis of Bachelier \cite{Bachelier 00} which was completed at the end of the 
XIXth century, more than a hundred years ago. The ideas of Bachelier were further 
carried out by many scholars \cite{Cootner 64,Merton 90}.

The price dynamics in financial markets can be mapped onto a 
random walk whose properties are studied in continuous, rather than
discrete, time \cite{Merton 90}. Here, we shall present this
mapping, pioneered by Bachelier \cite{Bachelier 00}, 
in a rather general way. It is worth mentioning that
this approach is related to that of Clark \cite{Clark 73} and to the
introductory notes in Parkinson's paper \cite{parkinson77}. 
As a further comment, this is a purely phenomenological approach.
No specific assumption on the rationality or
the behaviour of market agents is taken or even necessary.
In particular, it is not necessary to assume the
validity of the efficient market hypothesis \cite{fama70,fama91}.
Nonetheless, as shown below, a phenomenological model can
be useful in order to empirically corroborate or falsify the consequences
of behavioural or other assumptions on markets. 
Moreover, the model itself can be corroborated or falsified by empirical data.

As a matter of fact, there are various ways in which
random walk can be embedded in continuous time. Here, we shall base our approach on the
so-called continuous-time random walk in which time intervals between successive steps are random variables, as discussed by Montroll and Weiss \cite{montroll65}.

Let $S(t)$ denote the price of an asset or the value of an index at time
$t$. In a real market, prices are fixed when buy orders are matched with
sell orders and a 
transaction (trade) occurs. Returns rather than 
prices are more convenient. For this reason, we shall take
into account 
the variable $x(t) = \log S(t)$, that is the logarithm of the price.
Indeed, for a small price variation $\Delta S = S(t_{i+1}) - S(t_{i})$, the 
return $r = \Delta S/S(t_{i})$ and the logarithmic return 
$r_{log} = log[S(t_{i+1}) / S(t_{i})]$ virtually coincide. 

As we mentioned before, 
in financial markets, not only prices can be modelled as random variables, 
but also waiting times between two consecutive 
transactions vary in a stochastic fashion. 
Therefore, the time series $\{ x(t_i) \}$
is characterised by $\varphi(\xi, \tau)$, the 
{\em joint probability density}
of log-returns $\xi_{i} = x(t_{i+1}) - x(t_{i})$ and of waiting times 
$\tau_i = t_{i+1} - t_{i}$. The joint density
satisfies the normalization condition 
$\int \int d \xi d \tau \varphi (\xi, \tau) = 1$. Both $\xi_i$ and 
$\tau_i$ are assumed to be independent and identically distributed (i.i.d.) random
variables.

Montroll and Weiss \cite{montroll65} have shown that the Fourier-Laplace
transform
of $p(x,t)$, the probability density function, {\em pdf}, of
finding 
the value $x$ of the price logarithm (which is the diffusing quantity in 
our case) at time $t$, is:
\begin{equation}
\label{montroll}
\widetilde{\widehat p}(\kappa, s) = {1-\widetilde \psi(s) \over s}\,
{1 \over 1- \widetilde{\widehat \varphi}(\kappa, s)}\,,
\end{equation}
where 
\begin{equation}
\label{transform}
\widetilde{\widehat p}(\kappa, s) = 
\int_{0}^{+ \infty} dt \; \int_{- \infty}^{+ \infty} dx \, 
\e^{\,\ds -st+i \kappa x} \, p(x,t)\,,
\end{equation}
and $\psi(\tau) = \int d \xi \, \varphi(\xi, \tau)$ 
is the waiting time pdf.

The space-time version of eq. (\ref{montroll}) can be derived by probabilistic considerations \cite{mainardi00}. The following integral equation gives the probability density, $p(x,t)$, for the walker being in position $x$ at time $t$, conditioned by the fact that it was in position $x=0$ at time $t=0$:
\begin{equation}
\label{masterequation}
p(x,t) =  \delta (x)\, \Psi(t) +
   \int_0^t \, 
 \int_{-\infty}^{+\infty}  \varphi(x-x',t-t')\, p(x',t')\, dt'\,dx',
\end{equation}
where $\Psi(\tau)$ is the so-called survival function. $\Psi(\tau)$ is related to the marginal waiting-time probability density $\psi(\tau)$. The survival function $\Psi(\tau)$ is:
\begin{equation}
\label{survival}
\Psi(\tau) = 1 - \int_{0}^{\tau} \psi (\tau') \, d \tau' = \int_{\tau}^{\infty} \psi (\tau') \, d \tau'.
\end{equation}

The CTRW model can be useful in applications such as speculative option pricing by Monte Carlo simulations or portfolio selection. This will be the subject of a forthcoming paper. Here, it is more interesting to discuss the relation of this formalism to compound Poisson processes. Indeed, compound Poisson processes are an instance of continuous-time random walks in which waiting times and log-returns are independent random variables; moreover, one assumes that the marginal waiting-time density $\psi(t)$ is an exponential
density:
\begin{equation}
\label{exponentialdensity}
\psi( \tau ) = \mu \e^{-\mu \tau}.
\end{equation}
Therefore, the probability $P(n,t)$ of getting $n$ log-price jumps up to time $t$
is given by the Poisson distribution:
\begin{equation}
\label{Poissondistribution}
P(n,t) = \frac{( \mu t )^n}{n!} \e^{- \mu t},
\end{equation}
that is the jump point process is a Poisson process. 
The log-price $x(t)$ at time $t$ is:
\begin{equation}
\label{logpriceprocess}
x(t) = \sum_{i=1}^{n(t)} \xi_i.
\end{equation}
where, as above, $n(t)$ is the number of jumps occurred up to time $t$.
Let $\lambda(\xi)$ denote the marginal log-return density, then the solution of
eq. (\ref{masterequation}) is:
\begin{equation}
\label{compoundpoisson}
p(x,t) = \sum_{n=0}^{\infty} \frac{( \mu t)^n}{n!} \e^{- \mu t} \lambda_n (x),
\end{equation}
where $\lambda_n$ is the $n$-fold convolution of the density $\lambda$. Eq. (\ref{compoundpoisson}) can be also derived by purely probabilistic consideration. The interested reader can find more information on a generalization of this case in a recent paper of our group \cite{scalas04}. An important property of CTRWs is that log-returns and waiting times are independent and identically distributed random variables. Still,
there can be a dependence between the two random variables. If they are independent, as in the case of compound Poisson processes, the 
joint pdf $\varphi(\xi,\tau)$ is given by the product of the two marginal densities:
\begin{equation}
\label{independence}
\varphi(\xi, \tau) = \lambda(\xi) \psi(\tau);
\end{equation}
if they are not independent, then, according to the definition of conditional probability, one has:
\begin{equation}
\label{dependence}
\varphi(\xi, \tau) = \lambda(\xi) \psi(\tau|\xi) = \lambda(\xi|\tau) \psi(\tau),
\end{equation}
where $\psi(\tau|\xi)$ and $\lambda(\xi|\tau)$ are conditional probability densities. 
Note, however, that autoregressive conditional duration models introduce a dependence between waiting times and this feature cannot be captured by the above formalism, as 
waiting times are assumed to be i.i.d. random variables
(see also ref. \cite{salmon01}).

\section{Empirical evidence}

\subsection{The data set}

The data set consists of nearly 800,000 prices $S(t_i)$ and times of execution $t_i$ obtained from the TAQ database of the NYSE. These data were appropriately filtered in order to remove misprints in prices and times of execution and correspond to the high-frequency trades registered at NYSE in October 1999, for the 30 stocks of the Dow Jones Industrial Average Index, namely, at that time: AA, ALD, AXP, BA, C, CAT, CHV, DD, DIS, EK, GE, GM, GT, HWP, IBM, IP, JNJ, JPM, KO, MCD, MMM, MO, MRK, PG, S, T, UK, UTX, WMT, XON. The choice of one month of high-frequency data was a trade off between the necessity of managing enough data for significant statistical analyses and and, on the other hand,
the goal of minimizing the effect of external economic fluctuations. The reader can determine the company to which the above symbols correspond just by consulting the NYSE web pages ({\tt www.nyse.com}).

In order to roughly evidence intraday patterns \cite{dacorogna01}, the data set has been divided into three daily periods: morning (from 9:00 to 10:59), midday (from 11:00
to 13:59) and afternoon (from 14:00 to 17:00). In Table 1, the number of trades for each
daily period is given as a function of the stock.

\begin{table}[h]
\begin{center}
\begin{tabular}{|c|c|c|c|}
\hline
Stock&n1 (9:00-10:59)&n2 (11:00-13:59)&n3 (14:00-17:00)\\
\hline\hline
AA&4098&5662&5298\\
\hline
ALD&5248&7367&6504\\
\hline
AXP&9054&12267&12988\\
\hline
BA&5058&7080&6717\\
\hline
C&15628&21578&18541\\
\hline
CAT&3596&5361&4790\\
\hline
CHV&4973&6608&5591\\
\hline
DD&5284&7363&6913\\
\hline
DIS&7160&10504&9182\\
\hline
EK&3218&4433&4174\\
\hline
GE&16063&20214&19372\\
\hline
GM&16134&4340&6173\\
\hline
GT&3124&4105&3968\\
\hline
HWP&10278&14095&12062\\
\hline
IBM&12534&22668&16633\\
\hline
IP&4358&6263&5590\\
\hline
JNJ&6693&9856&8644\\
\hline
JPM&6410&7704&7991\\
\hline
KO&8511&12437&10575\\
\hline
MCD&5641&7729&6895\\
\hline
MMM&3578&5398&4996\\
\hline
MO&9680&14565&11852\\
\hline
MRK&9222&13462&11587\\
\hline
PG&6809&9598&8482\\
\hline
S&4694&5838&5319\\
\hline
T&12291&18598&14391\\
\hline
UK&2738&3305&3208\\
\hline
UTX&3745&5765&5249\\
\hline
WMT&8344&12446&10256\\
\hline
XON&9321&11669&10838\\
\hline
\end{tabular}
\end{center}
\caption{For each daily period, the total number of corrected monthly trades is given for each DJIA stock traded in October 1999.}\label{Period}
\end{table}

\subsection{Empirical analysis}

In Fig. 1, the waiting-time complementary cumulative distribution function (or survival function) $\Psi(\tau) = 1- \int_{0}^{\tau} \psi(t') dt'$ is plotted for three different periods of the day and for the GE time series of October 1999. In the above formula, $\psi(\tau)$ represents the marginal waiting-time probability density function. $\Psi(\tau)$ gives the
probability that the waiting time between two consecutive trades is greater than the given $\tau$.
The lines are the corresponding standard exponential complementary cumulative distribution functions: 
\begin{equation}
\label{expsurv}
\Psi(\tau) = \exp(-\tau/\tau_0), 
\end{equation}
where $\tau_0$ is the empirical average waiting time. An eye inspection already shows the deviation of the real distribution from the exponential distribution. This fact is corroborated by the Anderson-Darling test \cite{stephens74}. According to this test, for a large number of samples, one has to compute
the following statistics, after ordering the samples $\tau_i$ in ascending order:
\begin{equation}
\label{AD1}
A^2 = [-m -S] \cdot \left[ 1 + \frac{0.6}{m} \right],
\end{equation}
where $m$ is the total number of samples and $S$ is
\begin{equation}
\label{AD2}
S= \sum_{i=1}^{m} \frac{(2i-1)}{m} \{ \ln [F(\tau_i)] + \ln [1 - F(\tau_{m+1-i})] \},
\end{equation}
where $F$ is the survival function. In order to test the exponential distribution,
one must insert in the above formula the survival function
(\ref{expsurv}) with $\tau_0$ taken from the empirical estimates in Table 2. In the case of GE (Fig. 1),
the Anderson-Darling (AD) $A^2$ values for the three daily periods are, respectively, 352, 285, and 446. Therefore, the null hypothesis of exponential distribution can be rejected at the 1 \% significance level as the limit value is 1.957.

In Table 2, the values of the AD $A^2$ statistics are given for all the 30 DJIA stocks traded in October 1999. In all these cases the null hypothesis of exponentiality can be rejected at the 1 \% significance level.

It is interesting to observe that the average waiting time is sytematically and significantly larger at midday than in the morning or in the afternoon. This results points to a variable NYSE trade activity and is in agreement with previously
reported behaviour in stock markets \cite{raberto99,plerou00,bertram04}. This fact has
a {\em biological} explanation. Around midday the activity is slower as traders move from
their desks to eat. In fact,
as will be seen, these intra-day variations in trading activity
may also account for the reported anomaly in the distribution of
waiting times.

\subsection{Independent results corroborating this study}

Our study demonstrates that the marginal density for waiting times is definitely not an exponential function. After the publication of our paper series \cite{scalas00,mainardi00,gorenflo01}, different waiting-time scales have been investigated in different markets by various authors. All these empirical analyses corroborate the waiting-time anomalous behaviour. A study on the waiting times in a contemporary FOREX exchange and in the XIXth century Irish stock market was presented by Sabatelli {\it et al.} \cite{sabatelli02}. They were able to fit the Irish data by means of a Mittag-Leffler function as we did before in a paper on the waiting-time marginal distribution in the German-bund future market \cite{mainardi00}. Kyungsik Kim and Seong-Min Yoon studied the tick dynamical behavior of the bond futures in Korean Futures Exchange (KOFEX) market and found that the survival probability displays a stretched-exponential form \cite{kim03}. Moreover, just to stress the relevance of non-exponential waiting times, a power-law distribution has been
recently detected by T. Kaizoji and M. Kaizoji in analyzing the calm time interval of price changes in the Japanese market \cite{kaizoji03}.

\begin{figure}
\begin{center}
\mbox{\epsfig{file=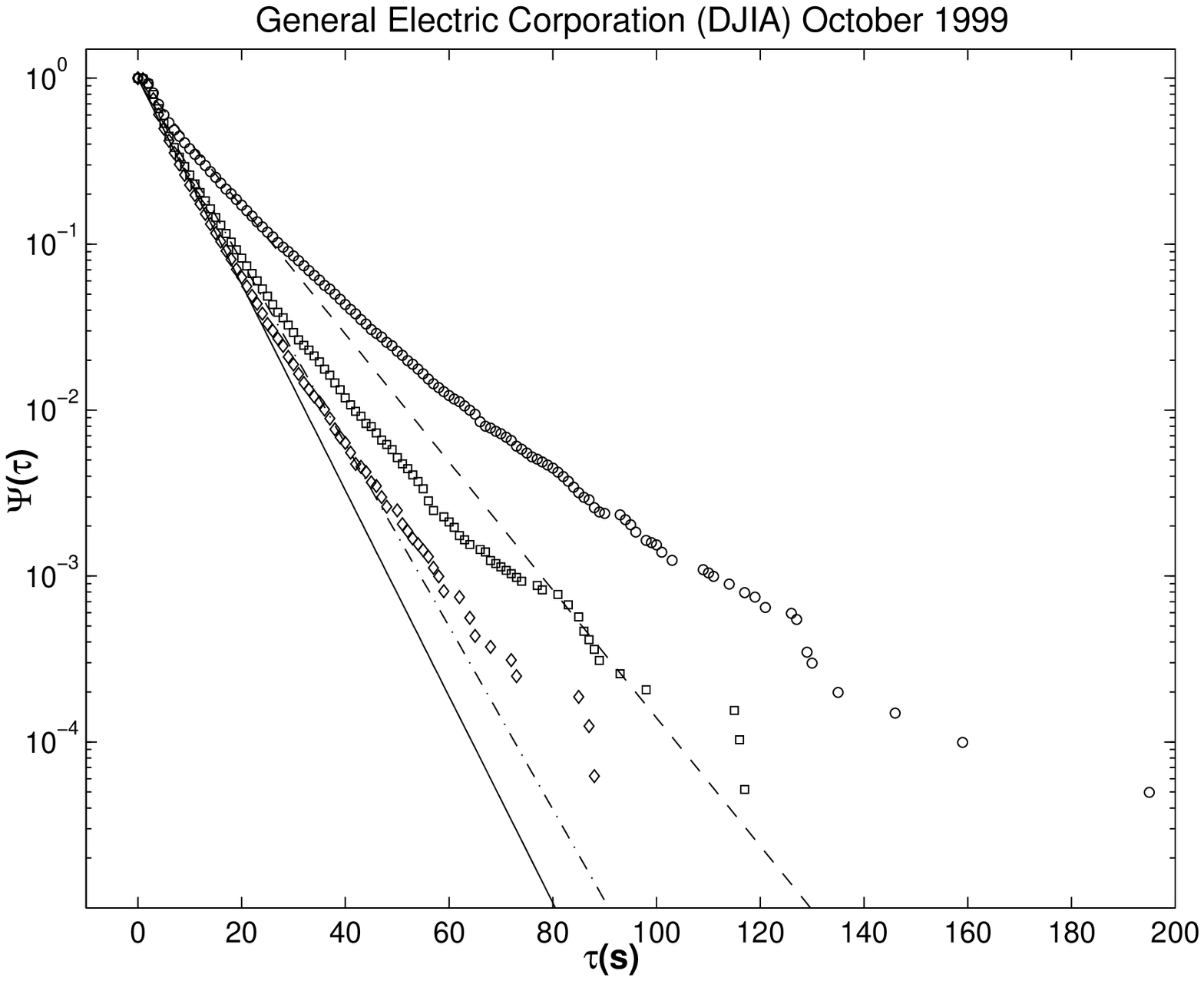,width=5in,angle=0}}
\end{center}
\caption{Waiting-time complementary cumulative distribution function $\Psi(\tau)$ for GE trades quoted at NYSE in October 1999. Open diamonds represent $\Psi(\tau)$ for the morning hours (9:00 -- 10:59). There were 16063 trades in this period in October 1999. The solid line is the corresponding standard exponential complementary cumulative distribution function with $\tau_0 = 7.0$ s. Open circles represent $\Psi(\tau)$ for the period around midday (11:00 -- 13:59). There were 20214 trades in this period in October 1999. The dashed line is the corresponding standard exponential complementary cumulative distribution function with $\tau_0 = 11.3$ s. Open squares represent $\Psi(\tau)$ for the afternoon hours (14:00 --  17:00). There were 19372 trades in this period in October 1999. The dash-dotted line is the corresponding standard exponential complementary cumulative distribution function with $\tau_0 = 7.9$ s. The day was divided into three periods to evidence seasonalities (see text for explanation).}
\label{f.1}
\end{figure}

\begin{table}
\begin{center}
\begin{tabular}{|c|c|c|c||c|c|c|}
\hline
Stock&$\tau_{0}^{mo} (s)
$&$\tau_{0}^{mi} (s)
$&$\tau_{0}^{af} (s) $&$A^{2}(mo)$&$A^{2}(mi)$&$A^{2}(af)$ \\
\hline\hline
AA&27.1&40.0&28.8&29.2&66.0&44.8\\
\hline
ALD&21.2&30.8&23.4&21.8&55.5&33.8\\
\hline
AXP&11.8&18.5&11.7&81.7&102.5&130.7\\
\hline
BA&22.0&32.0&22.6&17.4&20.2&21.2\\
\hline
C&7.1&10.5&8.2&252.2&142.8&210.7\\
\hline
CAT&29.2&42.4&31.6&72.3&128.7&64.6\\
\hline
CHV&22.1&34.3&27.1&104.4&121.5&64.9\\
\hline
DD&20.3&30.8&22.1&22.9&44.3&36.1\\
\hline
DIS&15.2&20.8&16.6&53.4&53.4&74.7\\
\hline
EK&34.1&51.2&36.3&24.8&34.8&44.3\\
\hline
GE&7.0&11.3&7.9&351.9&284.7&445.6\\
\hline
GM&24.6&36.6&27.0&22.4&60.8&40.9\\
\hline
GT&34.3&55.5&37.9&73.7&95.7&54.1\\
\hline
HWP&10.4&16.1&12.7&94.8&77.8&100.8\\
\hline
IBM&8.9&10.0&9.2&409.6&472.5&489.5\\
\hline
IP&24.8&36.3&27.0&25.0&37.2&19.4\\
\hline
JNJ&16.1&23.0&17.7&30.4&35.6&38.0\\
\hline
JPM&17.0&29.5&19.0&33.0&85.2&85.8\\
\hline
KO&12.9&18.3&14.4&44.5&37.8&44.1\\
\hline
MCD&19.4&29.3&22.1&40.9&72.7&44.1\\
\hline
MMM&30.1&42.0&30.4&80.1&86.8&37.5\\
\hline
MO&11.4&15.6&12.9&74.2&89.0&75.2\\
\hline
MRK&11.7&16.8&13.2&133.1&136.0&189.8\\
\hline
PG&16.2&23.6&17.9&43.5&37.2&48.8\\
\hline
S&23.4&38.8&28.6&40.1&23.0&41.6\\
\hline
T&8.8&12.2&10.6&193.2&179.1&208.9\\
\hline
UK&40.4&69.1&46.7&33.8&72.4&47.2\\
\hline
UTX&28.5&39.3&29.0&33.7&62.9&58.0\\
\hline
WMT&12.5&18.2&14.9&105.2&110.6&139.1\\
\hline
XON&12.0&19.6&14.1&104.8&121.4&129.0\\
\hline
\end{tabular}
\end{center}
\caption{For each daily period, the table gives the values of the empirical average waiting time $\tau_{0}$ and the AD statistics $A^{2}$
\cite{stephens74}.}\label{ADtest}
\end{table}

\section{Discussion and conclusions}

Why should we care about these empirical findings on the waiting-time distribution? This has to do both with the market price formation mechanisms and with the bid-ask process. {\it A priori}, one could argue that there is no strong reason for independent market investors to place buy and sell orders in a time-correlated way. This argument would lead one to expect a Poisson process. If price formation were a simple thinning of the bid-ask process, then exponential waiting times should be expected between consecutive trades as well \cite{cox79}. Eventually, even if empirical analyses should show that time correlations are already present at the bid-ask level, it would be interesting to understand why they are there. In other words, the empirical results on the survival probability set limits on statistical market models for price formation. A possibly correlated result has been recently obtained by Fabrizio Lillo and Doyne Farmer, who find that the signs of orders in the London Stock Exchange obey a long-memory process \cite{lillo03} as well as by Jean Philippe Bouchaud and coworkers \cite{bouchaud04}. Further studies on market microstructure will be necessary to clarify this point.

However, it is possible to offer a simple explanation of the anomalous behaviour in terms of exponential mixtures due to variable activity during the trading day.

Let us introduce a toy model of variable activity during a trading day. The trading day can be divided into $N$
subintervals where waiting times follow an exponential distribution with different average waiting times $\tau_{0,1}, \ldots, \tau_{0,N}$. Just recalling that the rate is the inverse of the average waiting time: $\mu_i = 1/\tau_{0,i}$,
one has that the survival function is given by:
\begin{equation}
\label{hyperexp}
\Psi (\tau) = \sum_{i}^{N} a_i \e^{-\mu_i \tau},
\end{equation}
where $a_i$ are suitable weights whose sum $\sum_{i=1}^{N} a_i$ must be 1, to fulfill the condition $\Psi(0) = 1$. This sum of exponential components is itself non-exponential. For illustrative purposes, in Fig. 2, the reader can find the comparison between eq. (\ref{hyperexp}) and simulated data in which the day
had been divided into 10 intervals of equal weight. In each interval the average waiting time between trades was a constant and the waiting times followed an exponential distribution. The value of the constant increased from 10 to 50 seconds in the first five intervals and then decreased from 40 to 5 seconds in the last five intervals, so that the sequence of waiting times (in seconds: 10,20,30,40,50,40,30,20,10,5) is a rough representation of the activity in a real financial market. The open circles are the survival function of the Monte Carlo simulation, the solid line represents the single exponential fit of the survival function, whereas, the crosses are values of the survival function computed according to eq. (\ref{hyperexp}) with $a_i = 1/10$. Even if for long waiting times, the tail of the distribution is again exponential with rate $\mu_i = 1/5$, the exponential mixture can describe deviations from the single exponential law for short and intermediate waiting times.

The probability density corresponding to eq. (\ref{hyperexp}) can be formally written in the following way:
\begin{equation}
\label{hyperpdf}
\psi(\tau) = \sum_{i=1}^{N} \mu_i \e^{-\mu_i \tau}
\end{equation}
Eq. (\ref{hyperpdf}) can be readily extended to a continuous spectrum of rates, $g(\mu)$:
\begin{equation}
\label{hyperpdfcont}
\psi(\tau) = \int_{0}^{\infty} \mu \e^{-\mu \tau} g(\mu) \, d \mu,
\end{equation}
where the condition $\int g(\mu) \, d \mu =1$ must hold.
Indeed, the integral equation (\ref{hyperpdfcont}) reduces to eq. (\ref{hyperpdf}) if $g(\mu)$ has the following form:
\begin{equation}
\label{discrspectr}
g(\mu) = \sum_{i=1}^{N} a_i \delta(\mu - \mu_i),
\end{equation}
where $\delta(\bullet)$ is Dirac's generalized function and $\sum_{i=1}^{N} a_i = 1$. 

In conclusion, we have shown that, in October 1999, waiting times between consecutive trades in the 30 NYSE DJIA stocks were non-exponentially distributed. We have summarized other recent results pointing to the same conclusions for different markets. We have argued that this fact has implications for market microstructural models that should be able to reproduce such a non-exponential behaviour to be realistic. Finally, we have offered a possible explanation in terms of variable trading activity during the day.  

\begin{figure}
\begin{center}
\mbox{\epsfig{file=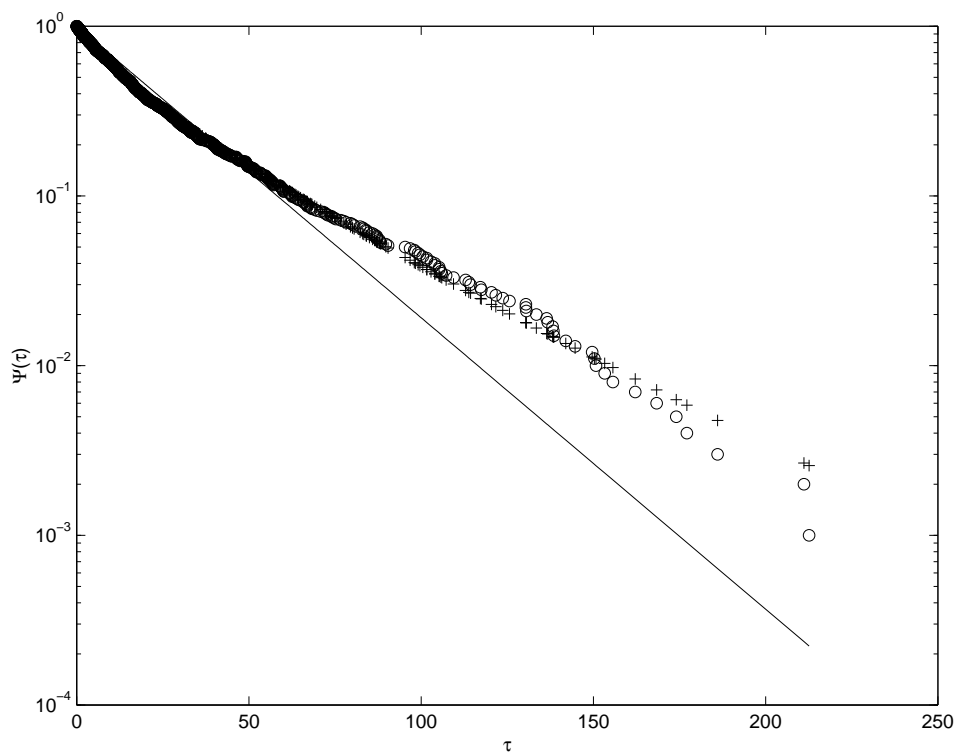,width=5in,angle=0}}
\end{center}
\caption{Waiting-time complementary cumulative distribution function $\Psi(\tau)$ for simulated data (open circles)
compared to a simple exponential fit (solid line) and to a mixture of exponentials (crosses). See text for details.}
\label{f.2}
\end{figure}

\section*{Acknowledgements}

We would like to acknowledge useful discussions with Sergio Focardi of The Intertek Group. This work was supported by grants from the Italian
M.I.U.R. Project COFIN 2003  ``Order and Chaos in nonlinear extended
systems: coherent structures, weak stochasticity and anomalous transport'' and by the Italian M.I.U.R. F.I.S.R. Project ``Ultra-high frequency dynamics of financial markets''.

\end{document}